\documentclass[10pt,aps,pra,twocolumn,superscriptaddress,floatfix,nofootinbib]{revtex4-2}
\makeatletter
\def\@bibdataout@aps{
 \immediate\write\@bibdataout{
  @CONTROL{
   apsrev41Control,author="08",editor="1",pages="0",title="0",year="1",eprint="1"
  }
 }
 \if@filesw
  \immediate\write\@auxout{\string\citation{apsrev41Control}}
 \fi
}
\makeatother

\usepackage{lipsum}
\usepackage{amsmath}
\usepackage{amsthm}
\usepackage{amsfonts, amssymb, bbm, braket, accents}
\usepackage{graphicx}
\usepackage{wrapfig}
\usepackage[usenames,dvipsnames]{color}
\usepackage[makeroom]{cancel}
\usepackage{soul}
\usepackage{tikz} 
\usetikzlibrary{shapes.arrows} 
\usetikzlibrary{shapes.geometric}
\usetikzlibrary{calc} 
\usepackage[makeroom]{cancel}
\usepackage{verbatim}
\usepackage{pifont}
\usepackage[mathscr]{euscript}

\usepackage{booktabs}
\usepackage{xcolor}
\usepackage{siunitx}
\usepackage{colortbl}

\definecolor{cellmin}{rgb}{1,1,1}
\definecolor{cellmax}{rgb}{0.25,1,0.25}
\definecolor{cellpurp}{rgb}{0.8, 0.6, 0.8}
\newcommand{\cellcolorbymax}[1]{\cellcolor{cellmax!#1!cellmin}}

\newcommand{\cellcolorpurp}[1]{\cellcolor{cellpurp!#1!cellpurp}}

\newcommand{\brak}[1]{\left\langle #1\right\rangle}

\newcommand{\expt}[1]{\langle {#1} \rangle}

\newcommand{\probP}{\text{I\kern-0.15em P}}
\newcommand{\probE}{\text{I\kern-0.15em E}}

\usepackage[breaklinks=true]{hyperref}
\hypersetup{
  colorlinks   = true,
  urlcolor     = blue,
  linkcolor    = blue,
  citecolor   = red
}

\usepackage[capitalise]{cleveref}
\crefformat{equation}{Eq.~(#2#1#3)}
\crefformat{section}{Sec.~#2#1#3}
\Crefformat{equation}{Equation~(#2#1#3)}
\crefformat{figure}{Fig.~#2#1#3}
\crefrangeformat{equation}{Eqs.~#3(#1)#4--#5(#2)#6}
\Crefformat{section}{Section~#2#1#3}

\begin{document}
\title{Sub-universal variational circuits for combinatorial optimization problems}

\author{Gal Weitz}
\altaffiliation{These two authors contributed equally.}
\affiliation{Department of Physics, University of Colorado Boulder, Boulder, CO 80309, USA}

\author{Lirandë Pira}
\altaffiliation{These two authors contributed equally.}
\affiliation{University of Technology Sydney,
		Centre for Quantum Software and Information,
		Ultimo NSW 2007, Australia}
\email{lpira@nus.edu.sg}

\author{Chris Ferrie}
\affiliation{University of Technology Sydney,
		Centre for Quantum Software and Information,
		Ultimo NSW 2007, Australia}

\author{Joshua Combes}
\affiliation{Department of Electrical, Computer and Energy Engineering, University of Colorado Boulder, Boulder, Colorado 80309, USA}

\date{\today}

\begin{abstract}
Quantum variational circuits have gained significant attention due to their applications in the quantum approximate optimization algorithm and quantum machine learning research. This work introduces a novel class of classical probabilistic circuits designed for generating approximate solutions to combinatorial optimization problems constructed using two-bit stochastic matrices. Through a numerical study, we investigate the performance of our proposed variational circuits in solving the Max-Cut problem on various graphs of increasing sizes. Our classical algorithm demonstrates improved performance for several graph types to the quantum approximate optimization algorithm. Our findings suggest that evaluating the performance of quantum variational circuits against variational circuits with sub-universal gate sets is a valuable benchmark for identifying areas where quantum variational circuits can excel.
\end{abstract}

\maketitle

\section{Introduction}\label{sec:intro}
There is much interest in constructing parameterized quantum circuits as variational ansätze to solve mathematical problems~\cite{varqalg}. Such circuits can find applications in quantum chemistry, for example, using variational quantum eigensolvers~\cite{VQE2014} or quantum machine learning~\cite{bilkis_2022}. This class of circuits also includes the quantum approximate optimization algorithm (QAOA) of \citet{farhi2014quantum}, which has received significant attention. Its purpose is to efficiently approximate the global optima of constraint satisfaction problems (CSPs) in combinatorial optimization.

A popular constraint satisfaction problem that is relevant in physics is called Max-Cut. The solution to Max-Cut can be used to find the minimum energy of the Ising Hamiltonian~\cite{MohseniMcMahon2022}. The Max-Cut problem is defined on a graph as follows. If the vertices of a graph can take one of two labels, the objective of Max-Cut is to maximize the number of edges with opposing labels. The constraint comes from the topology of the graph. If we have $n$ vertices, the optimum assignment is somewhere among the $2^n$ combinations. Finding the exact minimum (or maximum) of Max-Cut is considered an NP-complete problem; see, e.g., Appendix A2.2 of~\cite{garey1979computers}. While there are no current algorithms (beyond brute force searching) that guarantee an exact solution to Max-Cut, there are classical and quantum approximation techniques that produce ``good-enough'' solutions in polynomial time.

\begin{figure}[t!]
\includegraphics[width=0.60\linewidth]{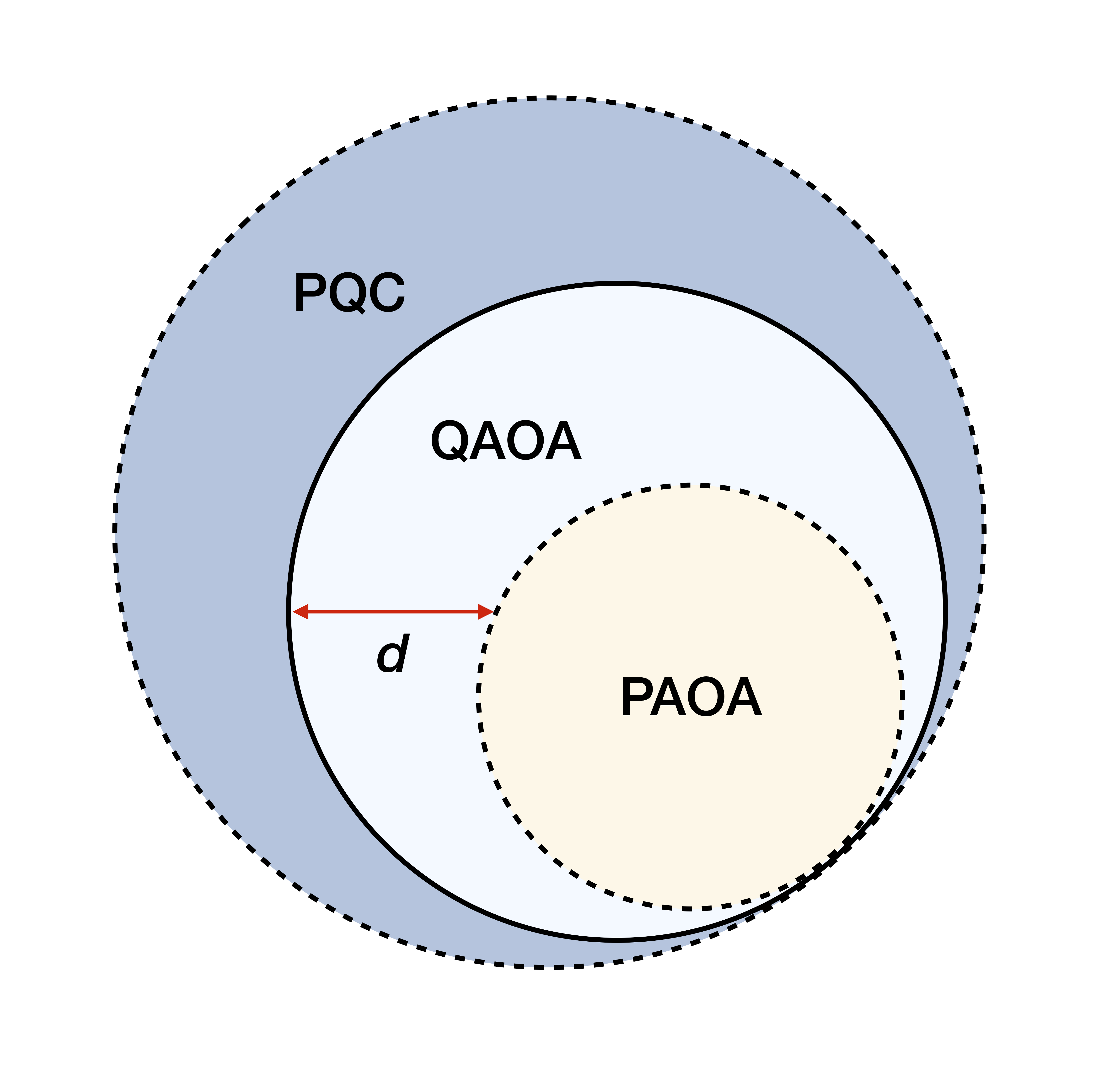}
\caption{Perspective on the relationship between three model classes considered in this work. Parameterized quantum circuits (PQCs) are the paradigm that represents the largest class of parameterized quantum algorithms. QAOA (Farhi et al. \cite{farhi2014quantum}) is a popular example. PAOA represents a class of probabilistic methods that are entirely classical. The ``distance'' between classical and quantum methods we consider to be in the practical sense --- i.e., do they achieve comparable performance \textit{in practice} where near-term quantum computers are expected to be used.
}\label{figcircle}
\end{figure}

The simplest solution is random guessing, which would ``cut'' half of the edges in the graph, on average. The fraction of the number of cut edges to the optimal solution is known as the approximation ratio. The gold standard of classical techniques is the Goemans-Williamson algorithm, which guarantees a ratio of 0.8785 ~\cite{gw}.

On the quantum side, Farhi et al. \cite{farhi2014quantum} proved that for a depth one ($p=1$) quantum circuit on 3-regular, triangle-free graphs, the QAOA guarantees a solution with an approximation ratio of 0.6924. It was also shown that as $p \rightarrow \infty$, the QAOA will always find the true optimal solution. Recent experiments have implemented the QAOA on larger numbers of qubits (see, e.g., Refs.~\cite{otterbach_unsupervised_2017,PaganoMonroe2020,Harrigan_2021,EbadiLukin2022}). However, these demonstrations solve Max-Cut on small graphs, and the time to solution is slow relative to real-world applications. Furthermore, the Goemans-Williamson algorithm has an almost optimal approximation ratio and, with small modifications, can efficiently solve problems on sparse graphs with 20 million vertices on a laptop~\cite{Yurtsever_SDP_2021} and thus impacts the onset of quantum advantage~\cite{StilckGarcia2021}. Given these factors, one might question the relevance of using QAOA. Some possible responses are (i) variational approaches like QAOA might process broader applicability than the semidefinite programming (SDP) relaxation, and (ii) QAOA could outperform the Goemans-Williamson algorithm on extremely large problems as it is a hardware-based solution. This raises the question of what the appropriate classical protocol to compare is.

In this article, we introduce the \textit{probabilistic} approximate optimization algorithm (PAOA), a classical probabilistic variational circuit inspired by the QAOA, that could be implemented as a hardware-based solution. In numerical experiments, we have found that PAOA can achieve performance comparable, and in some cases superior, to QAOA. The protocol we suggest provides a fairer comparison to classical techniques than the typical approach of comparing to random guessing. This provides a way to benchmark QAOA's performance in a way closer in spirit than the bound provided by the Goemans-Williamson algorithm and other excellent approaches like that of Refs.~ \cite{hastings_classical_2019,BapatJordan2021}. This paper intends not to compete with the Goemans-Williamson algorithm but to provide a compelling alternative to quantum solutions. 

At the core of the PAOA protocol lies probabilistic bits or ``$p$-bits.'' These are intermediate between standard classical bits and qubits. While qubits can be prepared in a superposition state of ``0'' and ``1'', p-bits can only be prepared in classical mixtures of 0 and 1. Interestingly, p-bits can be physically implemented using modern technology ~\cite{pbits_datta_2017,2019pbits,2019hamiltonianpbit}. Interestingly, there has been recent and renewed interest in p-bits see e.g. Refs.\cite{Aadit2022,MisraBland2023,Chowdhury2023,coles_thermodynamic_2023,kobayashi_cmos_2023,Lou2023,li_stochastic_2023,JohnSunAppenzeller2024,WangChenShen2024}. If these p-circuits (probabilistic circuits) can be implemented using current technology, yet yield similar results to that achieved using current quantum circuits, is there a real quantum advantage? The envisioned relationship between these classes is illustrated in Figure \ref{figcircle}.

In \Cref{sec:csp_maxcut}, we briefly review constraint satisfaction problems and Max-Cut. In \Cref{sec:qaoa}, we summarize QAOA. In \Cref{sec:paoa}, we introduce a class of classical variational circuits that can solve some constraint satisfaction problems. We compare QAOA to our classical variational circuits numerically in \Cref{sec:results}. Our numerics indicate that a low-depth classical variational algorithm has comparable performance to a quantum algorithm. We conclude with a summary and discussion of some open questions in \Cref{sec:conclusion}.

\section{Constraint satisfaction problems and Max-Cut}\label{sec:csp_maxcut}

A constraint satisfaction problem is specified by a set of items $\{z\}$ and $m$ constraints. Each constraint involves a subset of the items. The computational task is to find a combination of items that maximizes the number of satisfied constraints. For each constraint $a\in[m]$ and each string in a given CSP, we define,
\begin{equation}
    C_a(z)=
    \begin{cases}
       1 & \text{if $z$ satisfies constraint a}, \\
       0 & \text{otherwise} \\
    \end{cases}.
\end{equation}
Hence, the goal is to maximize the cost function $C(z)$ over the set $\{z\}$, where,
\begin{equation}
    C(z)=\sum_{a=1}^{m}C_a(z).
\end{equation}

Max-Cut is a CSP defined on a graph $G(z,E)$.  We represent the set of vertices as $n$-bit strings, i.e., $z=z_nz_{n-1}...z_2z_1$, and the set of edges $\langle ij \rangle \in E$. We consider a simple graph $G(z,E)$ and denote the number of edges $|E|$. For each vertex $i$, we assign a label, denoted as $z_i\in\{0,1\}$. Consequently, the $n$-bit string $z$ becomes a string of ones and zeros of length $n$, denoted as $z\in\{0,1\}^n$. Thus, there are $2^n$ distinct states spanning the set $\{0,1\}^n$, each representing a unique assignment of the vertices. The goal of Max-Cut is to find the maximum number of edges whose vertices on each end have different labels over the set $\{0,1\}^n$. Equivalently, we find the string $z$ that maximizes,
\begin{equation}\label{eq:classical_cost}
    C(z)=\sum_{\brak{ij}\in E}C_{\brak{ij}}(z),
\end{equation}
where,
\begin{equation}\label{eq:cost}
    C_{\brak{ij}}(z)=\frac{1}{2}\big(1-(-1)^{z_i}(-1)^{z_j}\big).
\end{equation}

\subsection{Quantum reformulation of  Max-Cut objective}\label{sec:qmaxcut}

The maximization of $C(z)$ corresponds to finding the maximum energy of a Hamiltonian $C$, an operator in the $2^n \times 2^n$ dimensional Hilbert space with basis vectors $\ket{z}\in\mathbb{C}^n$, defined by the eigenvalue equation
\begin{equation}
   C\ket{z}=C(z)\ket{z},
\end{equation}
where $C(z)$ is defined in \cref{eq:classical_cost}. For a simple graph $G(z,E)$, each vertex $i$ is associated with a spin state $\ket{z_i}$, where $z_i\in\{0,1\}$ and $Z_i\ket{z_i}=(-1)^{z_i}\ket{z_i}$. The state associated with the entire graph is $\ket{z}=\otimes^n_{i=1}\ket{z_i}$ and  $Z_i\ket{z}=(-1)^{z_i} \ket{z}$ where $Z_i$ is identity on all spins except the $i^{th}$ spin. The quantum representation of the cost operator of Max-Cut is
\begin{equation}\label{eq:q_cost_0}
    C=\sum_{\brak{ij}\in E}C_{\brak{ij}}, \,\,\, {\rm where}\quad 
    C_{\brak{ij}}=-\frac{1}{2}(I-Z_iZ_j).
\end{equation}

\section{Quantum Approximate Optimization Algorithm}\label{sec:qaoa}
Here we summarize QAOA, following the presentation of \citet{farhi2014quantum}. The objective of QAOA is to find an $n$-bit string $z$ that approximately minimizes the cost $C$. Using $C$ from Eq. \eqref{eq:q_cost_0}, we define the unitary cost operator $U(C,\gamma)$,
\begin{equation}\label{eq:costop}
    U(C,\gamma)=e^{-i\gamma C}=\prod_{\brak{jk}\in E}e^{-i\frac{\gamma}{2}(Z_jZ_k-I)},
\end{equation}
where $\gamma\in[0,2\pi)$ applies a phase to pairs of bits according to the cost function. Now we also define the operator $B$,
\begin{equation}
    B=\sum_{k=1}^{n}X_k,
\end{equation}
with $X_k$ the single qubit Pauli $X$ operator operating on the qubit corresponding to the $k^{th}$ bit in $z$. We define the ``mixer'' unitary operator with $\beta\in[0,\pi)$,
\begin{equation}
    U(B,\beta)=e^{-i\beta B}=\prod_{k=1}^{n}e^{-i\beta X_k}.
\end{equation}
This unitary operator drives transitions between bit-strings within a superimposed state \cite{Harrigan_2021}. 

QAOA involves a sequential application of $U(C,\gamma)$ and $U(B,\beta)$ to an initially uniform superposition of computational basis states.  Let $ \ket{+_n}:=H^{\otimes n}\ket{0}^\otimes n$ denote the uniform superposition of all possible states, where $H$ is the Hadamard gate.
The \textit{depth number} of the circuit, $p$, is an integer number that counts sequential applications of the two unitaries,
\begin{equation}
    U(\boldsymbol{\gamma},\boldsymbol{\beta})=\prod_{k=1}^{p}U(B,\beta_k)U(C,\gamma_k),
\end{equation}
with a total of $2p$ angles defined via  $\boldsymbol{\gamma}:=(\gamma_1,...,\gamma_p)$ and $\boldsymbol{\beta}:=(\beta_1,...,\beta_p)$. Thus, the state after one application of the circuit is
\begin{equation}\label{eq:gamma_beta_state}
    \ket{\boldsymbol{\gamma},\boldsymbol{\beta}} = U(\boldsymbol{\gamma},\boldsymbol{\beta})\ket{+_n},
\end{equation}
and the expectation of the cost operator in the final state is
\begin{equation}\label{eq:qaoa_exp}
    \langle C \rangle=\bra{\boldsymbol{\gamma},\boldsymbol{\beta}}C\ket{\boldsymbol{\gamma},\boldsymbol{\beta}}.
\end{equation}
It is the vector of parameters $\boldsymbol{\gamma},\boldsymbol{\beta}$ that we will optimize over. 

With an initial set of parameters, we conduct a repeated measurement of the state in Eq. \eqref{eq:gamma_beta_state}. Each measurement will have the superposition state collapse to one of the basis states with probability
\begin{equation}\label{eq:prob}
    \Pr(z)=|\langle z|\boldsymbol{\gamma},\boldsymbol{\beta}\rangle |^2.
\end{equation}
Given the binary string $ z $, we can efficiently compute the cut and record the corresponding cost $C(z)$. Repeating this process many times, we collect a sample of costs $\{C(z)\}$, from which we can get {\em estimates} for the expectation in Eq. \eqref{eq:qaoa_exp} and the minimum value of the sample
\begin{equation}
    C_{\min}=\min_z\{C(z)\}.
\end{equation}
We then define the ``true'' approximation ratio,
\begin{equation}\label{eq:true_ratio}
   R^*=\langle C \rangle/C_{\min}\,.
\end{equation}

Note that, in practice, $C_{\min}$ is rarely the absolute minimum cost achievable, as that would be the cost of the solution to the Max-Cut problem. For this reason we introduce $R$, which is an estimate of the ``true'' approximation ratio which we denote as
\begin{equation}\label{eq:ratio}
    R=\langle C \rangle_{\rm est}/C_{\min,\, \rm est}\,,
\end{equation}
where the subscript ``est'' indicates the quantity has been estimated from a finite amount of data or bit strings.

Once estimated, $R$ is treated as an objective function for a classical optimization algorithm. Hence, in an iterative fashion, QAOA will initialize the parameterized circuit with better problem-specific parameters, causing constructive and destructive interference to states that are better and worse for the problem, respectively. After optimization, the candidate optimal solution is 
\begin{equation}
    z^*=\mathop{\mathrm{argmin}}_z\{C(z)\}\,.
\end{equation} 

For $p=1$ on 3-regular graphs, we are guaranteed an approximation ratio that corresponds to $0.6924$ of the cost of the true optimal state \cite{farhi2014quantum}.  It was also shown in \cite{farhi2014quantum} that as $p$ goes to infinity, $C_{\min}$ corresponding to the solution is achieved, and the approximation ratio converges to $1$ as the distribution around the mean converges to the optimal cost of the problem.

\section{Classical probabilistic variational circuits}\label{sec:paoa}

In this section, we describe our classical probabilistic variational circuits. We begin by summarizing Markov chains and their application in reversible logic in \cref{sec:markovchain}. Then, we present our variational circuit ansatz in \cref{sec:ansatz} for solving the Max-Cut problem.

\begin{figure*}
    \centering
    \includegraphics[width=0.75\linewidth]{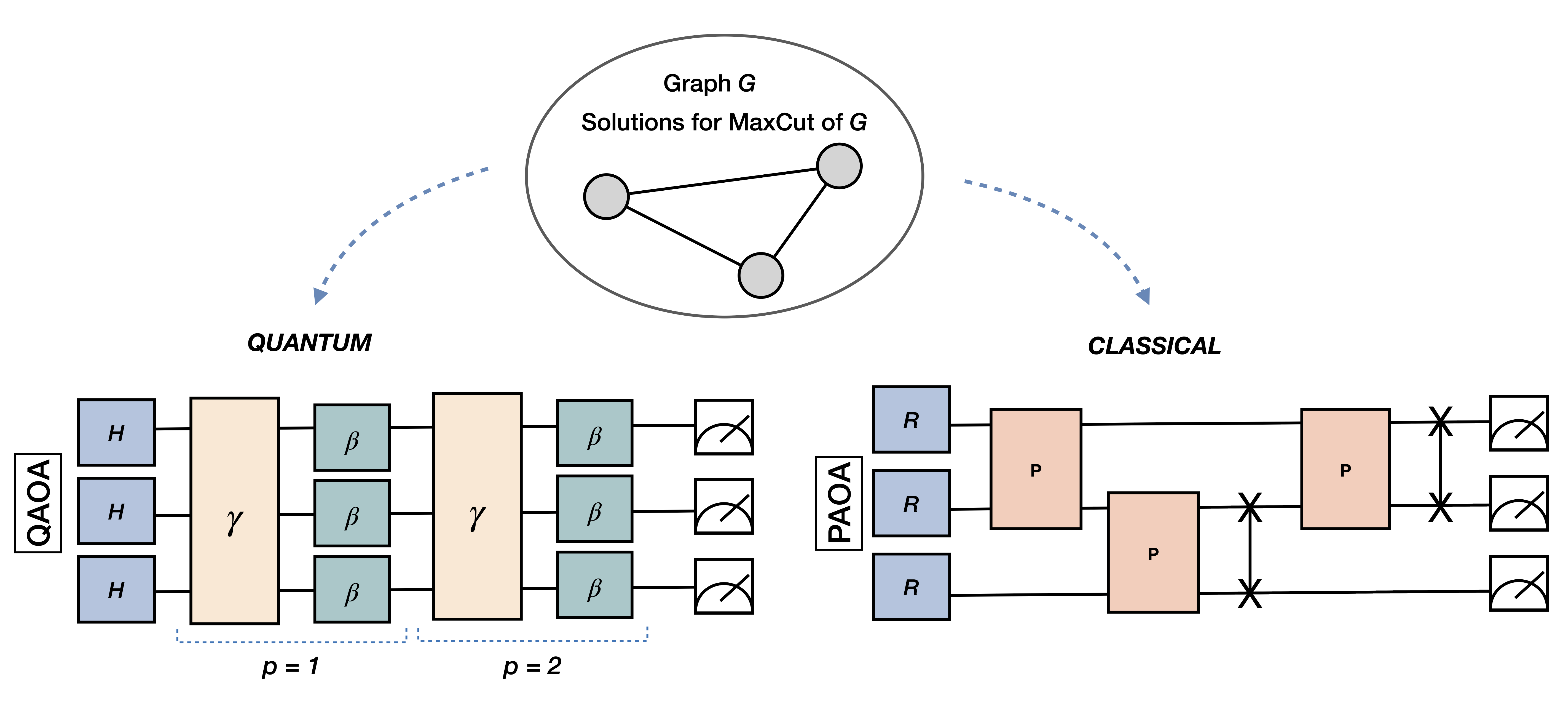}
    \caption{Illustrative example of the Max-Cut algorithms for a 3-node graph $G$ showing both QAOA and PAOA circuits. (Left) $H$ is the Hadamard gate and $p$ is the circuit depth (see Section \ref{sec:qaoa} for the in-depth explanation). (Right) $R$ denotes a random initial state, and $P$ represents the probabilistic gate ansatz we use (see Section \ref{sec:ansatz}). We assume it is possible to run the PAOA circuit directly on the graph as SWAPs are basically free in classical computation. Thus, the SWAPs depicted are virtual but allow us to map the linear circuit topology to the problem graph topology.
    }\label{fig:comp}
\end{figure*}

\subsection{Classical probabilistic circuits}\label{sec:markovchain}

Let us consider probability distributions over a classical bit, which have the zero entropy states,
\begin{equation}
    \ket{0} = (1,0)^\mathsf{T}, \quad     \ket{1} = (0,1)^\mathsf{T}.
\end{equation}
Although these are classical vectors, we use Dirac notation to make a stronger analogy with QAOA.  A linear combination of these vectors is a probabilistic-bit or p-bit,
\begin{equation}
    \ket{\psi} = (1-p)\ket{0} + p\ket{1} = (1-p, p)^\mathsf{T},
\end{equation}
where $p\in [0,1]$. 

The two possible logical operations on a single p-bit are identity and \textsc{NOT}, which correspond to the (Pauli) permutation matrices,
\begin{equation}\label{eq:1bitperm}
   I  = \begin{bmatrix}
        1 & 0 \\
        0 & 1 
    \end{bmatrix}, \quad {\rm and} \quad 
    X  = \begin{bmatrix}
        0 & 1 \\
        1 & 0 
    \end{bmatrix}.
\end{equation}
The convex hull of these two permutations gives rise to the bit-flip channel,
\begin{equation}\label{eq:doublestoch}
    \mathcal E_q \ket{\psi} 
    = (1-q) I\ket{\psi} + qX\ket{\psi} 
    = \begin{bmatrix}
        (1-q) & q \\
        q & (1-q) 
    \end{bmatrix}\ket{\psi} \, ,
\end{equation}
where $q\in [0,1]$.

For $n$ p-bits, the zero entropy states are the familiar computational basis states $|z\rangle$, $z\in\{0,1\}^n$, and the convex hull of the permutation matrices is the set of doubly stochastic matrices ~\cite{bhatia2013matrix,marshall1979inequalities}.

In general, the probability of moving from state $|j\rangle$ to state $|i\rangle$ is denoted $P_{ij}$, which need not be equal to $P_{ji}$. Such processes are physically motivated by, e.g., the probability of an excited state of an atom to decay to the ground state but not vice versa. A probability transition matrix $P$ for which $P_{ij}\neq P_{ji}$ for some $i,j$ is called a {\em stochastic matrix}. As the total probability of transitioning from state $|j\rangle$ to any other state in $\{0,1\}^n$ is 1, a stochastic matrix must obey
\begin{equation}\label{eq:condition}
    \sum_{i=1}^{2^n} P_{ij}=1,
\end{equation}
for all $j\in\{0,1\}^n$. In what follows, we will use stochastic matrices, rather than doubly stochastic matrices, to construct our circuit ansatz.

Since the Max-Cut problem is specified on a graph, binary strings will encode vertices. For each edge in the graph, the possible bit-string assignments for two vertices are $\{0,1\}^2=\{00,01,10,11\}$.
Thus, the general two-bit stochastic matrix describing the transitions between the states of the bits on edge $\langle k l \rangle$ is
\begin{equation}\label{eq:TPM}
        \boldsymbol{p}_{\langle k l \rangle} =
    \begin{bmatrix}
        P_{11} & P_{12} & P_{13} & P_{14}\\
        P_{21} & P_{22} & P_{23} & P_{24}\\
        P_{31} & P_{32} & P_{33} & P_{34}\\
        P_{41} & P_{42} & P_{43} & P_{44}
    \end{bmatrix}.
\end{equation}
For the task of optimization, $\boldsymbol{p}_{\langle k l \rangle}$ can be written as a 12-dimensional vector, noting the four linear constraints imposed by \cref{eq:condition}. 

Since the state would remain a ``classical'' mixture, it is obvious that the circuits created with these gates are not universal for quantum computing. Interestingly, they are not universal for classical reversible computation either --- it is known that three-bit gates (e.g., the Fredkin or Toffoli gate) are required to make classical reversible computing universal \cite{toffoli_reversible_1980, toffoli_reversible_1980_2}. Nevertheless, with these ``sub-universal'' resources, we can still solve the Max-Cut problem with high approximation ratios.

\subsection{PAOA}\label{sec:ansatz}

\begin{figure}[t!]
\includegraphics[width=0.99\columnwidth]{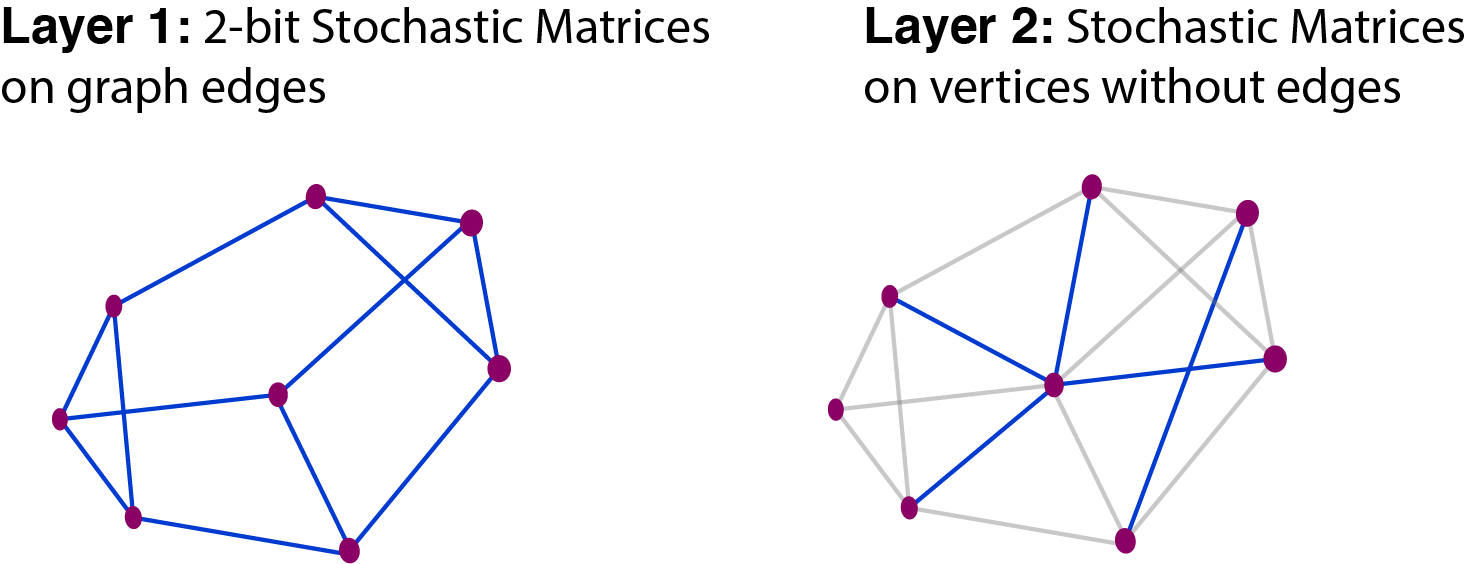}
\caption{(left) In the simple version of depth one ($p=1$) PAOA we optimize stochastic matrices on the problem graph. To make the probability distribution richer, we have to optimize either over higher-depth circuits or more than two-bit stochastic matrices. (right) The original problem graph is depicted in grey. A possible second layer of PAOA is depicted with the blue edges. }\label{fig:graph_anzatz_layers}
\end{figure}

In analogy with the parameterized quantum circuit of QAOA, here we propose a parameterized probabilistic circuit called the related protocol PAOA. Given a graph, a parameterized probabilistic circuit is constructed using independent stochastic matrices that act on some subset of the vertices, see \cref{fig:comp}. Generally, one could imagine applying one-, two-, and three-bit stochastic matrices to the vertex bits and iterating over layers as in QAOA.  The limiting case would be a single $n$ bit stochastic matrix, see \cref{eq:condition}, which is like a (classical) Boltzmann machine \cite{KORST1989331} or Ising machine~\cite{haribara2016coherent}.
However, in designing PAOA, we have found that a simple single-depth circuit of two-bit stochastic matrices is sufficient, and this has the benefit of reducing the dimensionality of the optimization space. Thus, for all of the variants we consider below, a depth 1 PAOA circuit creates a parameterized probability distribution over binary strings $z$ corresponding to the cut on the graph $G$ of the form
\begin{equation}
    \Pr(z|\boldsymbol{x}) = \prod_{\substack{\expt{kl}\in E\\k<l}}  \boldsymbol{p}_{\langle k l \rangle}(\boldsymbol{x}_{\expt{kl}})
\end{equation}
where $\boldsymbol{x}$ is the vector of all variational parameters for the graph $G$, and $\boldsymbol{x}_{\expt{kl}}$ are the variational parameters for the edge $\expt{kl}$. We will see that as our ansatz, only the distribution between two edges doing higher depth circuits doesn't add to the expressivity of our ansatz. Instead, we have to add different probability distributions to vertices that don't have edges, see \cref{fig:graph_anzatz_layers}.

In principle, we could use \cref{eq:TPM} to construct a variational ansatz for the edges of a graph, which would involve $12$ parameters for every edge in the graph. To reduce the number of parameters intelligibly, let's reexamine our objective \cref{eq:costop}. Recall the objective of Max-Cut: Find the arrangement of vertices that maximizes the number of edges with opposing bits on each end. 
Thus, we set transition probabilities to unwanted states, such as $\ket{01}\mapsto \ket{00}$, for example, to zero. This ensures maximum disagreement between edges. 

This motivates the following variational ansatz, which forms the basis of PAOA,
\begin{equation}\label{eq:TPOkl}
    \boldsymbol{p}_{\langle k l \rangle}^\textsc{paoa} =
    \begin{bmatrix}
        0 & 0 & 0 & 0\\
        p_1 & p_2 & p_3 & p_4\\
        1-p_1 & 1-p_2 & 1-p_3 & 1-p_4\\
        0 & 0 & 0 & 0
    \end{bmatrix}.
\end{equation}
The vector $\boldsymbol{p}$ only encodes four variational parameters per edge where $p_i\in [0,1]$. If the number of edges in the graph is $|E|$, PAOA has $4|E|$ parameters. This ansatz allows all of the four bitstrings between vertices to transition to the strings $\{01,10\}$ with unique probabilities.

In an effort to minimize the number of optimization parameters, we also construct Reduced PAOA, which sets all probabilities per edge to be equal, such that each edge is associated with a gate of the following form,
\begin{equation}\label{eq:TPOkl_reduced}
    \boldsymbol{p}_{\langle k l \rangle}^\textsc{r-paoa} =
    \begin{bmatrix}
        0 & 0 & 0 & 0\\
        p & p & p & p\\
        1-p & 1-p & 1-p & 1-p\\
        0 & 0 & 0 & 0
    \end{bmatrix},
\end{equation}
which amounts to $|E|$ parameters in total. Notice that if we multiply two of these stochastic matrices that $\boldsymbol{p}'\times \boldsymbol{p} = \boldsymbol{p}'$, so, in this case, larger depth circuits don't add expressivity to our ansatz.

The objective of the PAOA is to replicate QAOA classically (Figure \ref{fig:comp} provides an example of solution circuits using both methods). Given a graph, a parameterized probabilistic circuit is constructed using independent stochastic matrices for each edge. As one last ansatz, we define Min PAOA, which is parameterized as follows,
\begin{equation}\label{eq:TPOkl_reduced_min}
    \boldsymbol{p}_{\langle k l \rangle}^\textsc{min-paoa} =
    \begin{bmatrix}
        0 & 0 & 0 & 0\\
        p & q & 1-q & 1 - p\\
        1-p & 1-q & q & p\\
        0 & 0 & 0 & 0
    \end{bmatrix}
\end{equation}
which has \textit{the same} two parameters for every edge. Like QAOA, we allow this circuit to have multiple layers with different parameters. The form of this ansatz is chosen so that the transitions $\ket{00}\Leftrightarrow \ket{01}$ and $\ket{11}\Leftrightarrow \ket{10}$ are controlled by $p$, while $\ket{01}\Leftrightarrow \ket{10}$ are parameterized by $q$. Both Min PAOA and standard QAOA use $2p$ parameters.

Once the parameterized probabilistic circuit is defined, we follow the same protocol as in the QAOA. Namely, we conduct a sequence of experiments where the output of an experiment is the approximation ratio as defined in Eq. \eqref{eq:ratio}. After each experiment, an optimizer attempts to improve the choice of parameters such that the approximation ratio for the next experiment is larger than the previous one. We will discuss the details of our numerical experiments next.

\section{Numerical experiments}\label{sec:results}
\begin{figure*}[!ht]
\begin{center}
\begin{minipage}{0.18\textwidth}
$2$-regular graph
\end{minipage}
\begin{minipage}{0.18\textwidth}
$3$-regular graph
\end{minipage}
\begin{minipage}{0.18\textwidth}
Complete graph
 \end{minipage}
\begin{minipage}{0.18\textwidth}
Barabasi-Albert graph
\end{minipage}
\begin{minipage}{0.18\textwidth}
Erdős–Rényi graph
\end{minipage}

\begin{minipage}{0.18\textwidth}
  \includegraphics[width=\textwidth, height=0.8\textwidth ]{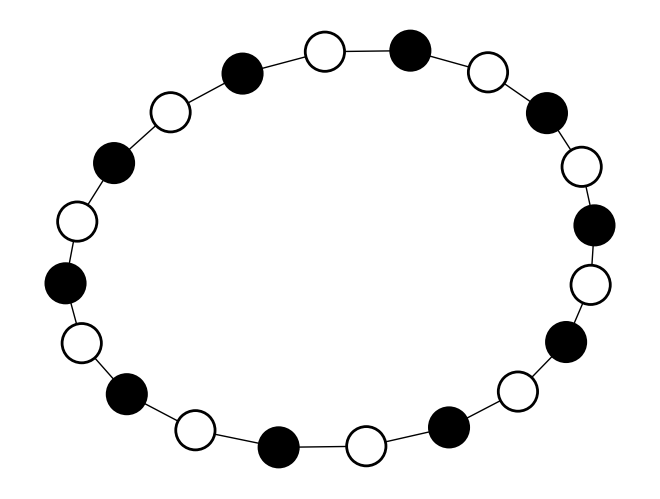}
\end{minipage}
\begin{minipage}{0.18\textwidth}
  \includegraphics[width=\textwidth, height=0.8\textwidth ]{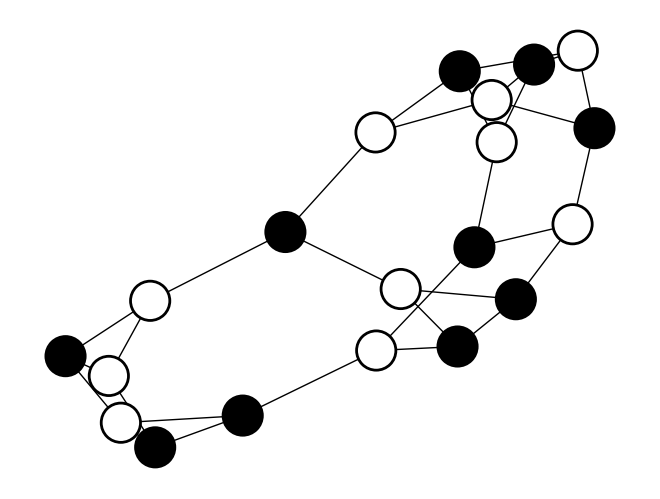}
\end{minipage}
\begin{minipage}{0.18\textwidth}
  \includegraphics[width=\textwidth, height=0.8\textwidth ]{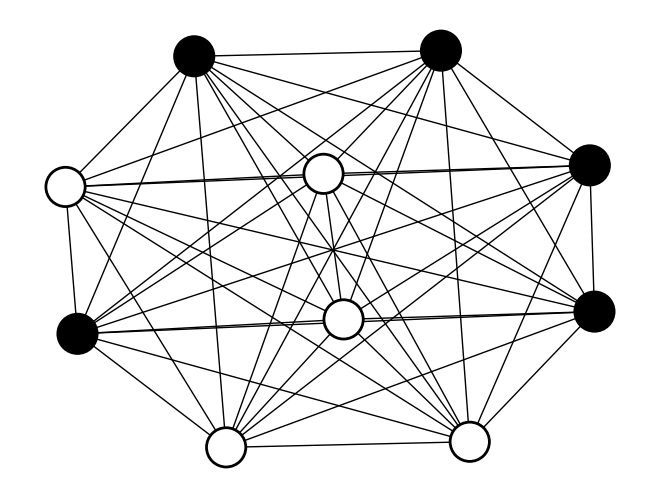}
 \end{minipage}
\begin{minipage}{0.18\textwidth}
   \includegraphics[width=\textwidth, height=0.8\textwidth ]{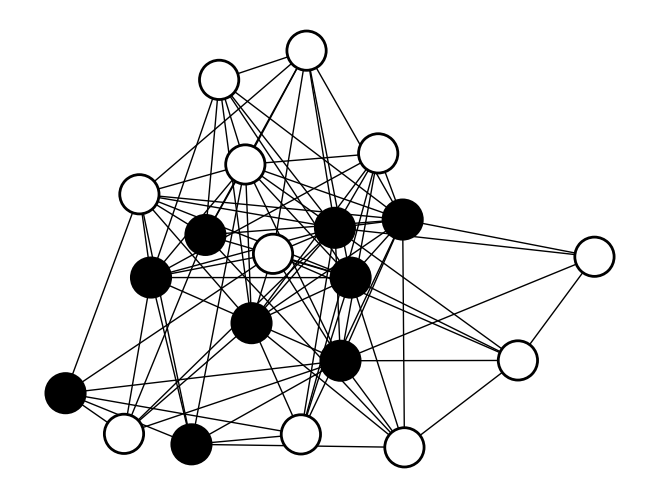}
 \end{minipage}
\begin{minipage}{0.18\textwidth}
    \includegraphics[width=\textwidth, height=0.8\textwidth ]{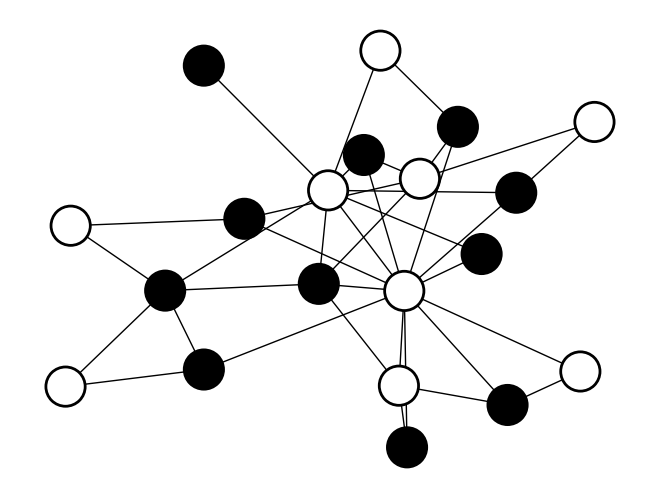}
 \end{minipage}
\end{center}
 \caption{ We consider running QAOA and variants of PAOA on these graphs to solve Max-Cut.
 }\label{fig:graphs}
\end{figure*}

In this section, we present numerical evidence for the effectiveness of PAOA on several graphs; see \cref{fig:graphs}. We start by considering small three-regular graphs with up to 10 edges and comparing the performance of QAOA to PAOA and show PAOA's performance is better than QAOA. 

As numerical simulations of large quantum systems are difficult, we then switch tact and simulate larger graphs. We consider several graph types and compare the performance of QAOA to PAOA on a graph with 20 edges. Then, we consider the scaling of the performance of PAOA to random guessing for graphs with 50 to 250 edges. For each considered graph, we run each Max-Cut optimization algorithm 100 times. The average and standard deviation of the cut sizes, as well as the estimated approximation ratio, will form the basis of our conclusions.

For the fairest comparison, an ``out of the box'' optimizer \cite{spsa_pypi} was used to train the QAOA and PAOA circuits. Each was given 100 iterations and 100 experimental runs per iteration. In no cases were the optimization algorithm's hyperparameters tuned (the defaults were used). The QAOA circuit was built and simulated using Qiskit \cite{qiskit}. The code to reproduce these results can be found at \cite{weitz_paoa_2023}.

\subsection{Performance as a function of graphs size for 3-regular graphs}

Regular, or $k$-regular graphs are those where each node has an equal ($k$) number of edges. For $k=0$, there are no edges. For $k=1$, the graph consists of disjoint pairs of nodes connected by a single edge, so the cut is trivial. Here, we will consider 3-regular graphs.

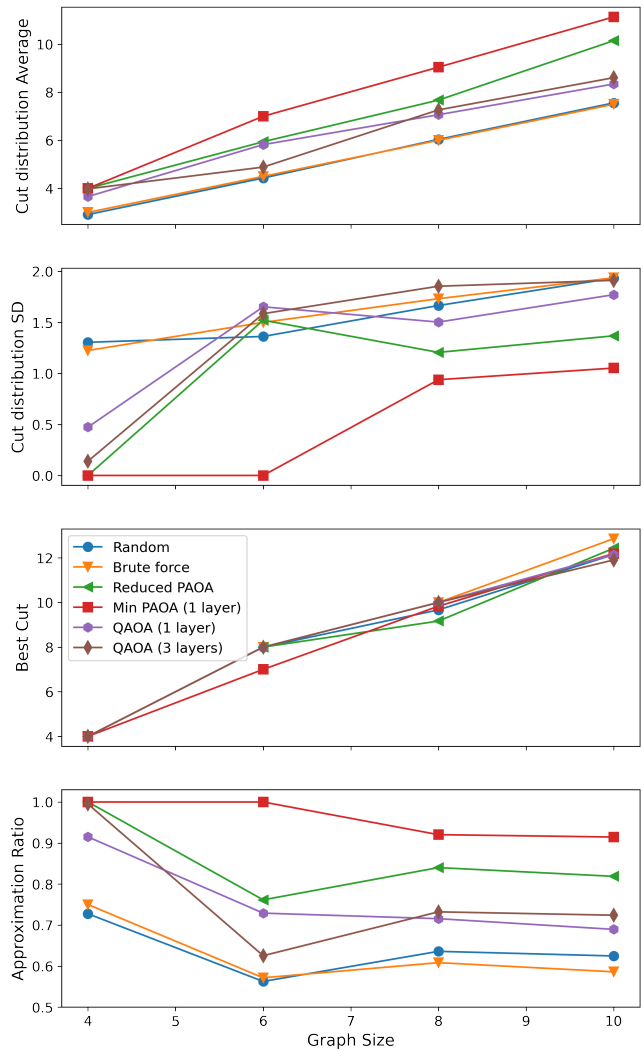
\begin{figure}[t!]
    \centering
\includegraphics[width=0.99\columnwidth]{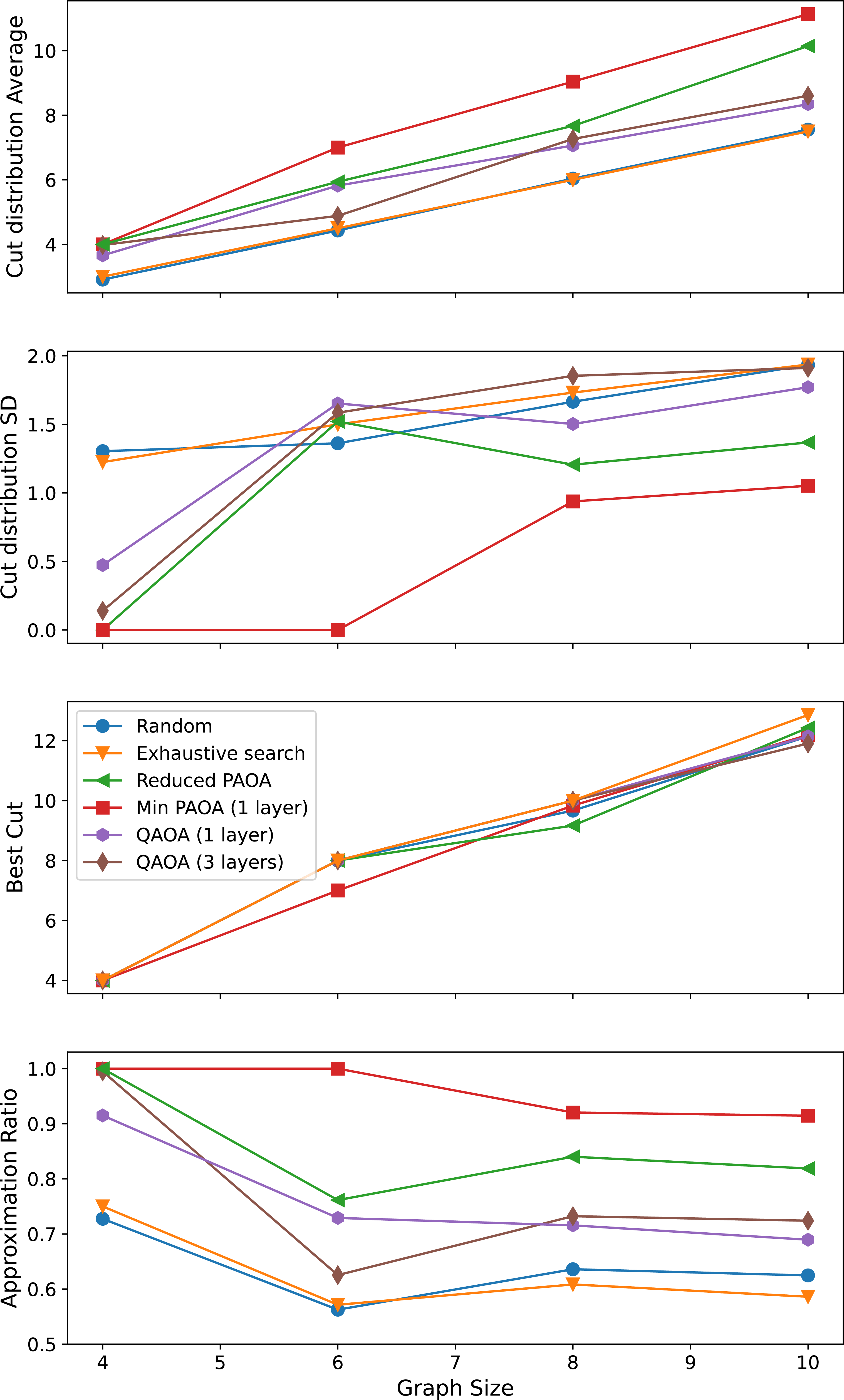}
    \begin{picture}(0,0)
        \put(-128,414){(a)}
        \put(-128,313){(b)}
        \put(-128,212){(c)}
        \put(-128,113){(d)}
    \end{picture}
\caption{A Comparison of the different protocols to solve Max-Cut on 3-regular graphs of increasing graph size, i.e., $|V|$. For each graph of size $|V|$, the protocols were given 100 iterations and 100 experimental runs per iteration. If there is more than one graph for a given size $|V|$, we average the performance metrics.
 }\label{fig:performance_comparison}
\end{figure}

\Cref{fig:performance_comparison} is a comparative analysis of the variants of PAOA, QAOA, random guessing, and exhaustive search methods for Max-Cut on 3-regular graphs of increasing size. Specifically we consider graphs with $|V|\in \{4,6,8,10\}$ where the number of corresponding 3-regular graphs are $\{1, 2, 6, 21 \}$~\cite[A005638]{oeis}. We run the different protocols on all possible graphs of size $|V|$ and average the results. For stronger guarantees about problem hardness, we could take an approach like that of Ref.~\cite{polloreno_qaoa_2023}, which explicitly looks for hard graphs.

It is worthwhile to pause here to explain the yet-to-be-defined methods of ``random guessing'' and ``exhaustive search''. In the random guessing method, we choose a random binary string (with replacement) and compute the cut. The difference between random guessing with and without replacement is asymptotically zero for large graphs. While the exhaustive search method systematically tries all binary strings and computes the cut. We make a distinction between an ``exhaustive search'' method and ``brute force,'' which seems to have connotations of post-selecting the best result of the exhaustive search method. So, for the remainder of this article, we will define ``brute force'' to be post-selecting the best result of the exhaustive search method.

We compare several protocols and several performance metrics as a function of graph size. The x-axis represents graph size, i.e., the order or number of vertices $|V|$, and the y-axis showcases the performance metrics. Across the graph sizes, PAOA consistently outperforms other methods,  with the exception of Best Cut, which the exhaustive search method obviously is the most performant. Moreover, PAOA's narrower distribution, denoted by a smaller standard deviation in \cref{fig:performance_comparison} (b), indicates its consistent and reliable performance. The fact that all methods get close to the exhaustive search method with respect to the best cut is not surprising, as the graphs are small, and our optimizer gets 100 shots. This also explains why the approximation ratios also look good. For this reason, we will study larger graphs in the latter part of this section.

Notice that we did not include the original PAOA ansatz, see \cref{eq:TPOkl}, because its performance is worse than ``Reduced PAOA'' and ``Min. PAOA'' which are \cref{eq:TPOkl_reduced,eq:TPOkl_reduced_min} respectively. We conjecture this is because of the curse of dimensionality in the original PAOA, which has $4|E|$. While Reduced and Min. PAOA has $|E|$ and $2$ parameters for each layer, respectively.

\subsection{Larger $2$ and $3$-regular graphs}\label{sec:reg_graphs}

In this section, we will focus on comparing the performance of QAOA and PAOA for larger graphs when $k=2$ or $k=3$. The maximum cut is trivial for $k=2$, where the graph (if fully connected) is a cycle. The maximum cut is clearly $n$, where $n$ is the (even) number of nodes. However, this example provides an interesting test case as we will soon see. We have also considered $k=3$, where the solution is non-trivial. The two example graphs corresponding to the data collected below are shown here. 

\begin{table}[h]
\centering
\caption{2-regular graph performance $(n=20)$}
\label{tab:2_performance}
\begin{tabular}{l c c c c} 
\toprule
Method & {Best} & {Average} & {SD} & {$R$} \\
\midrule
Brute Force        & \cellcolorpurp{75}   20    & \cellcolorpurp{75} 20         & \cellcolorpurp{75} 0      & \cellcolorpurp{75} 1 \\
Exhaustive search  & \cellcolorbymax{100} 20    & \cellcolorbymax{0} 10.00      & \cellcolorbymax{0} 2.24    & \cellcolorbymax{0} 0.50 \\
Random             & \cellcolorbymax{0}  16     & \cellcolorbymax{0} 10.16      & \cellcolorbymax{0} 2.34    & \cellcolorbymax{0} 0.64 \\
PAOA               & \cellcolorbymax{66} 18     & \cellcolorbymax{33} 14.58     & \cellcolorbymax{66} 1.53   & \cellcolorbymax{33} 0.81 \\
Reduced PAOA       & \cellcolorbymax{66} 18     & \cellcolorbymax{100} 16.72    & \cellcolorbymax{33} 1.54   & \cellcolorbymax{66} 0.93 \\
Min PAOA (1 layer) & \cellcolorbymax{66} 18     & \cellcolorbymax{66} 16.22     & \cellcolorbymax{100} 1.10  & \cellcolorbymax{100} 0.90 \\
Min PAOA (3 layers)& \cellcolorbymax{0} 16     & \cellcolorbymax{0} 11.24      & \cellcolorbymax{0} 2.17    & \cellcolorbymax{0} 0.70 \\
QAOA (1 layer)     & \cellcolorbymax{0} 16     & \cellcolorbymax{0} 10.76      & \cellcolorbymax{0} 1.97    & \cellcolorbymax{0} 0.67 \\
QAOA (3 layers)    & \cellcolorbymax{0} 16     & \cellcolorbymax{0} 11.12      & \cellcolorbymax{0} 1.95    & \cellcolorbymax{0} 0.69 \\
\bottomrule
\end{tabular}
\\ \vspace{6pt} 
\end{table}

\Cref{tab:2_performance} compares algorithm performance on a 2-regular graph with $n=20$ nodes across several metrics. Since the algorithms are probabilistic, each is run multiple times, producing a distribution of outcomes. We report:
Best -- the largest cut found in any trial,
Average -- the mean cut size over all trials,
SD -- the standard deviation of the cut size distribution, and
$R$ -- the approximation ratio see \cref{eq:ratio}.
Green shading highlights the three best results in each column, with opacity indicating rank: 100\% for the best, then 66\%  and 33\% opacity for second and third place.
In the first column, several results were equal, so they are given equal second place. 
The purple row at the top shows the brute-force method (exhaustive search plus postselection), where $R$ equals the ``true'' approximation ratio $R=R^*$.
Brute-force results are omitted from later tables, as they can be inferred from the exhaustive-search row.

One of the main takeaways from \cref{tab:2_performance} is the most performant protocol by the four comparison metrics is PAOA and variants. This is evident by the clustering of green on those rows of the table. This trend continues in the later tables, which examine different graphs. Surprisingly, this very simple graph seems to present a relative challenge for all algorithms. Of course, the exhaustive search algorithm will always yield the optimal cut. However, since it is searching over every possible cut, its performance on the other metrics is quite low. According to the approximation ratio cost function, Reduced PAOA is the best performer. However, a single layer of Min PAOA does quite well and is more ``reliable'' based on the standard deviation of the cut sizes it produces. QAOA and larger depth circuits seem to struggle with the chosen meta-heuristics. 

\begin{table}[h]
\centering
\caption{3-regular graph performance $(n=20)$}
\label{tab:3_performance}
\begin{tabular}{l c c c c}
\toprule
Method  & {Best} & {Average} & {SD} & {$R$} \\
\midrule
Exhaustive search  & \cellcolorbymax{100} 26     & \cellcolorbymax{0} 15.00   & \cellcolorbymax{0} 2.74     & \cellcolorbymax{0} 0.58 \\
Random             & \cellcolorbymax{0} 24       & \cellcolorbymax{0} 15.02   & \cellcolorbymax{0} 2.72     & \cellcolorbymax{0} 0.63 \\
PAOA               & \cellcolorbymax{0} 24       & \cellcolorbymax{33} 19.26  & \cellcolorbymax{100} 2.28   & \cellcolorbymax{33} 0.80 \\
Reduced PAOA       & \cellcolorbymax{100} 26     & \cellcolorbymax{66} 21.35  & \cellcolorbymax{0} 2.54     & \cellcolorbymax{66} 0.82 \\
Min PAOA           & \cellcolorbymax{100} 26     & \cellcolorbymax{100} 21.66 & \cellcolorbymax{0} 2.39     & \cellcolorbymax{100} 0.83 \\
Min PAOA (3 layers)& \cellcolorbymax{0} 22       & \cellcolorbymax{0} 15.99   & \cellcolorbymax{66} 2.30    & \cellcolorbymax{0} 0.73 \\
QAOA (1 layer)     & \cellcolorbymax{66.00} 25   & \cellcolorbymax{0} 17.62   & \cellcolorbymax{33} 2.32    & \cellcolorbymax{0} 0.70 \\
QAOA (3 layers)    & \cellcolorbymax{0} 24       & \cellcolorbymax{0} 17.75   & \cellcolorbymax{0} 2.81     & \cellcolorbymax{0} 0.74 \\
\bottomrule
\end{tabular}
\end{table}

In Table \ref{tab:3_performance}, the performance for a 3-regular graph with $n=20$ nodes is presented. In this case, both of the reduced parameter variants of PAOA find the solution with similar approximation ratios. While the approximation ratio of QAOA is better than random guessing, the standard deviation at deeper circuits is higher than PAOA.  

Over the course of many trials and observations not presented here, we note that either Min PAOA or Reduced PAOA are the best-performing algorithms for both $k=2$ and $k=3$ regular graphs. In either case, the number of gates (and hence the runtime) grows linearly with the number of nodes.

\begin{figure}[t!]
\includegraphics[width=0.99\columnwidth]{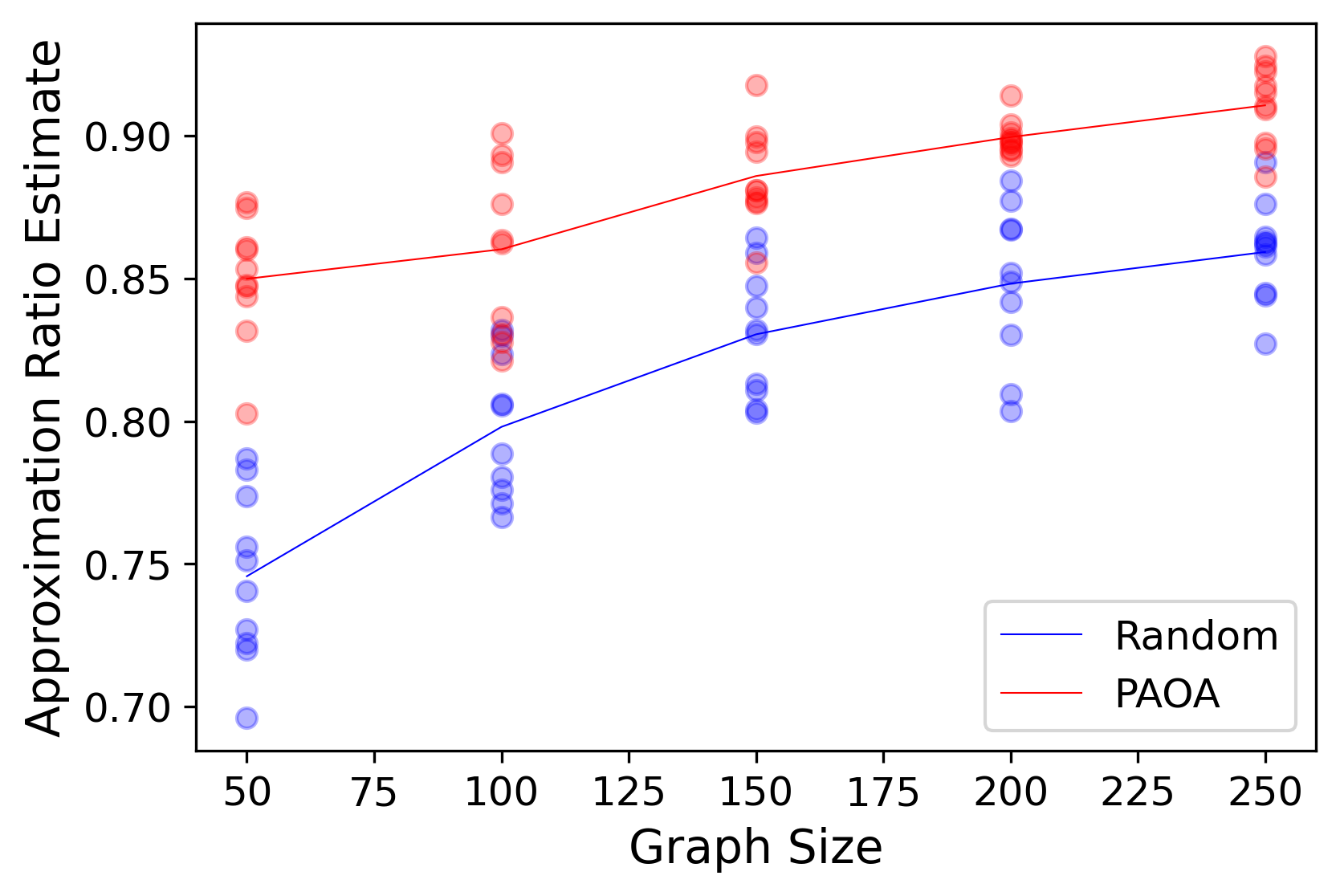}\\
\includegraphics[width=0.99\columnwidth]{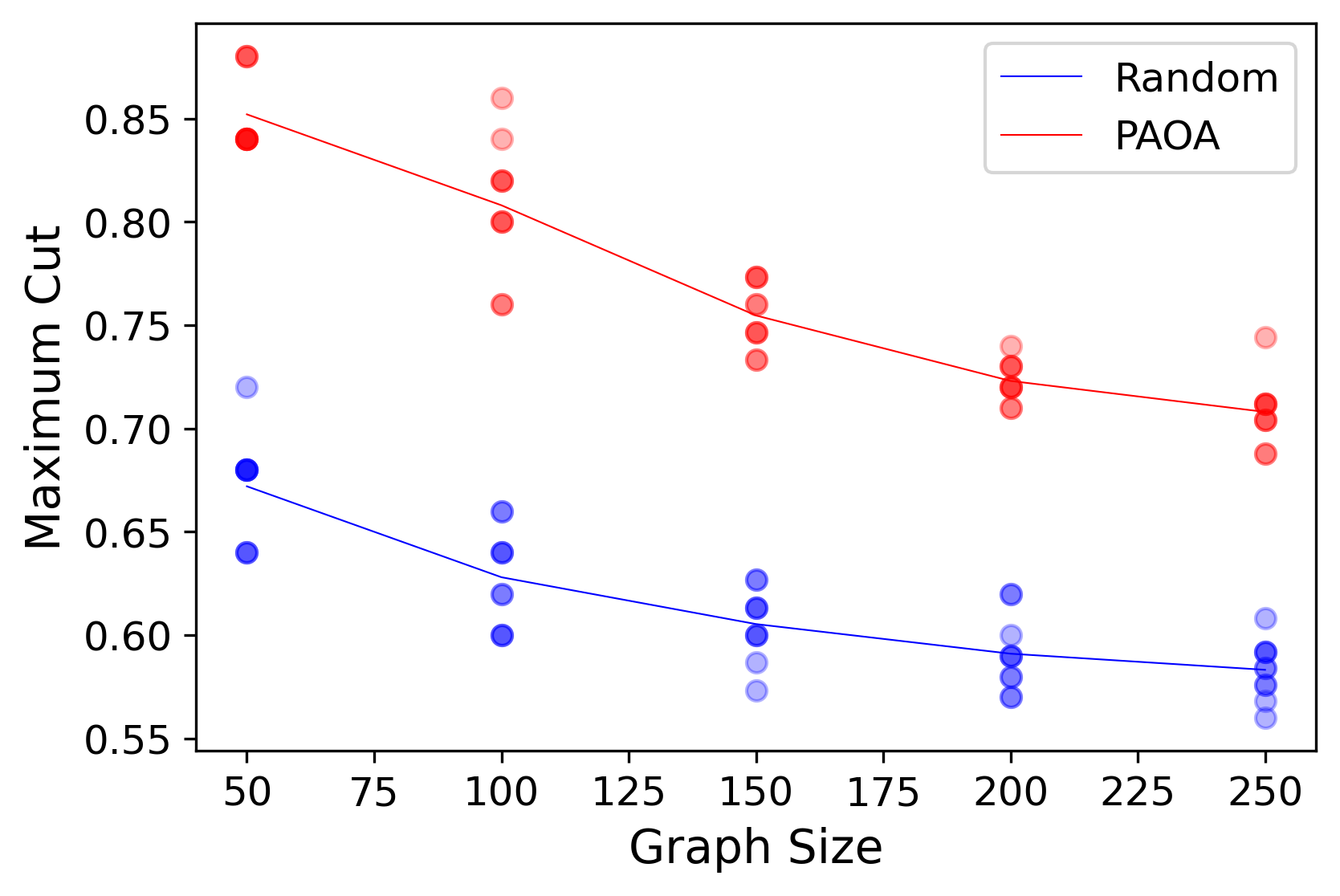}
\caption{A Comparison of Reduced PAOA and random guessing to solve max
cut on 2-regular graphs of increasing size. (Top) A comparison of the approximation ratios, and  (bottom) is the maximum cut normalized to the size of the graph, i.e. (the best cut)$/|V|$. Each algorithm was run ten times on a 2-regular graph. All data are presented along with the average (line).
 }\label{fig:paoa_ratio_2reg}
\end{figure}

Since QAOA was simulated classically, its space/time complexity was exponential in the number of graph edges --- thus, we were limited to the range below roughly 20 graph nodes. On the other hand, PAOA scales linearly with the number of edges. Since, as we will see in the following sections, Reduced PAOA is the best-performing algorithm, we tested its performance well beyond the 20 qubit limit. The results, summarized in Figure \ref{fig:paoa_ratio_2reg}, show that PAOA continues to outperform random guessing as the graph size increases. 

\subsection{Complete graphs}\label{sec:complete_g}

A complete graph is one in which every pair of nodes is connected by an edge; see \cref{fig:graphs}. By symmetry, any cut that equally partitions an even number of nodes will be the maximum. This was not difficult to find for any of the algorithms we considered. Note, however, that variational algorithms have runtimes that scale with the number of edges in a graph. While we did not intentionally benchmark the timing of each algorithm, complete graphs were certainly the most time-consuming to optimize.

The performance results of each algorithm on a complete graph with $n=10$ nodes are presented in Table \ref{tab:complete_performance}. Surprisingly, Min PAOA is nearly perfect, finding a set of variational parameters that produce a maximum cut deterministically for all practical purposes. The other variants of PAOA also performed extremely well.

\begin{table}[h]
\centering
\caption{Complete graph performance $(n=10)$}
\label{tab:complete_performance}
\begin{tabular}{l c c c c}
\toprule
Method & {Best} & {Average} & {SD} &  {$R$} \\
\midrule
Exhaustive search   & \cellcolorbymax{100} 25 & \cellcolorbymax{0} 22.50   & \cellcolorbymax{0} 3.35  & \cellcolorbymax{0} 0.90 \\
Random              & \cellcolorbymax{100} 25 & \cellcolorbymax{0} 22.74  & \cellcolorbymax{0} 3.86   & \cellcolorbymax{0} 0.91 \\
PAOA                & \cellcolorbymax{100} 25 & \cellcolorbymax{33} 24.14  & \cellcolorbymax{33} 1.30  & \cellcolorbymax{33} 0.97 \\
Reduced PAOA        & \cellcolorbymax{100} 25 & \cellcolorbymax{66} 24.52  & \cellcolorbymax{66} 0.92  & \cellcolorbymax{66} 0.98 \\
Min PAOA (1 layer)  & \cellcolorbymax{100} 25 & \cellcolorbymax{100} 24.99 & \cellcolorbymax{100} 0.10 & \cellcolorbymax{100} 1.00 \\
Min PAOA (3 layers) & \cellcolorbymax{100} 25 & \cellcolorbymax{0} 20.03   & \cellcolorbymax{0} 5.11   & \cellcolorbymax{0} 0.80 \\
QAOA (1 layer)      & \cellcolorbymax{100} 25 & \cellcolorbymax{0} 22.13   & \cellcolorbymax{0} 4.07   & \cellcolorbymax{0} 0.89 \\
QAOA (3 layers)     & \cellcolorbymax{100} 25 & \cellcolorbymax{0} 14.82   & \cellcolorbymax{0} 8.52   & \cellcolorbymax{0} 0.59 \\
\bottomrule
\end{tabular}
\end{table}

\subsection{Other random graphs}\label{sec:rand}

Lastly, we assessed the performance of the algorithms on some other popular random graphs from network theory. First, the Barabasi-Albert model is a technique for generating ``scale-free'' networks that mimic human-made networks such as the Internet. It uses a mechanism called \textit{preferential attachment} whereby new nodes are added to the graph with connections to existing nodes that exist probabilistically, weighted by how many connections each node currently has. The graph in \cref{fig:graphs} on the left was created starting with two connected nodes.

An Erdős–Rényi graph is one where each potential edge is added with some fixed and independent probability $p$. Here, and in \cref{fig:graphs}, we have chosen $p=\frac12$, which you can intuit as a complete graph with roughly half the edges removed.

In Table \ref{tab:ba_performance}, the performance of the algorithms is presented for a Barabasi-Albert graph on $n=20$ nodes. In this case, Reduced PAOA is a clear winner, with the other variants of PAOA falling slightly behind on all metrics. Again, QAOA seems to struggle to use the same meta-heuristics.

\begin{table}[h]
\centering
\caption{Barabasi-Albert graph performance $(n=20)$ }
\label{tab:ba_performance}
\begin{tabular}{l c c c c}
\toprule
Method & {Best} & {Average} & {SD}  & {$R$} \\
\midrule
Exhaustive search  & \cellcolorbymax{100} 29   & \cellcolorbymax{0} 18.00   & \cellcolorbymax{0} 3.00   & \cellcolorbymax{0} 0.62 \\
Random             & \cellcolorbymax{33} 27    & \cellcolorbymax{0} 18.39   & \cellcolorbymax{0} 3.04   & \cellcolorbymax{0} 0.68 \\
PAOA               & \cellcolorbymax{0} 26     & \cellcolorbymax{66} 22.68  & \cellcolorbymax{66} 2.22  & \cellcolorbymax{66} 0.87 \\
Reduced PAOA       & \cellcolorbymax{66} 28    & \cellcolorbymax{100} 24.91 & \cellcolorbymax{100} 2.04 & \cellcolorbymax{100} 0.89 \\
Min PAOA (1 layer) & \cellcolorbymax{66} 28    & \cellcolorbymax{33} 22.26  & \cellcolorbymax{0} 2.61   & \cellcolorbymax{0} 0.80 \\
Min PAOA (3 layers)& \cellcolorbymax{33} 27    & \cellcolorbymax{0} 21.69   & \cellcolorbymax{0} 3.03   & \cellcolorbymax{0} 0.80 \\
QAOA (1 layer)     & \cellcolorbymax{0} 23     & \cellcolorbymax{0} 18.59   & \cellcolorbymax{33} 2.41  & \cellcolorbymax{33} 0.81 \\
QAOA (3 layers)    & \cellcolorbymax{0} 26     & \cellcolorbymax{0} 18.79   & \cellcolorbymax{0} 2.70   & \cellcolorbymax{0} 0.72 \\
\bottomrule
\end{tabular}
\end{table}

The last detailed comparison appears in Table \ref{tab:er_performance}, where the performance of all algorithms is shown for an Erdős–Rényi graph on $n=20$ nodes. The results bear a resemblance to the complete graph, whereby all algorithms achieve a reasonably high approximation ratio. However, Reduced PAOA appears to be the most reliable, though it did not find the absolute optimal solution.

\begin{table}[h]
\centering
\caption{Erdős–Rényi graph performance $(n=20)$}
\label{tab:er_performance}
\begin{tabular}{l c c c c}
\toprule
Method & {Best} & {Average} & {SD} & {$R$} \\
\midrule
Exhaustive search  & \cellcolorbymax{100} 61   & \cellcolorbymax{0} 44.00   & \cellcolorbymax{0} 4.69     & \cellcolorbymax{0} 0.72 \\
Random             & \cellcolorbymax{0} 53     & \cellcolorbymax{0} 43.32   & \cellcolorbymax{0} 4.36    & \cellcolorbymax{0} 0.82 \\
PAOA               & \cellcolorbymax{100} 61   & \cellcolorbymax{66} 48.99  & \cellcolorbymax{0} 4.87     & \cellcolorbymax{0} 0.80 \\
Reduced PAOA       & \cellcolorbymax{66} 57    & \cellcolorbymax{100} 50.04 & \cellcolorbymax{100} 3.44   & \cellcolorbymax{100} 0.88 \\
Min PAOA (1 layer) & \cellcolorbymax{33} 55    & \cellcolorbymax{33} 45.81  & \cellcolorbymax{66} 3.53    & \cellcolorbymax{0} 0.83 \\
Min PAOA (3 layers)& \cellcolorbymax{33} 55    & \cellcolorbymax{0} 44.07   & \cellcolorbymax{0} 5.03     & \cellcolorbymax{0} 0.80 \\
QAOA (1 layer)     & \cellcolorbymax{0} 52     & \cellcolorbymax{0} 44.39   & \cellcolorbymax{33} 3.91    & \cellcolorbymax{66} 0.85 \\
QAOA (3 layers)    & \cellcolorbymax{0} 53     & \cellcolorbymax{0} 44.77  & \cellcolorbymax{0} 4.93     & \cellcolorbymax{33} 0.84 \\
\bottomrule
\end{tabular}
\end{table}

\begin{figure}[t!]
\includegraphics[width=0.99\columnwidth]{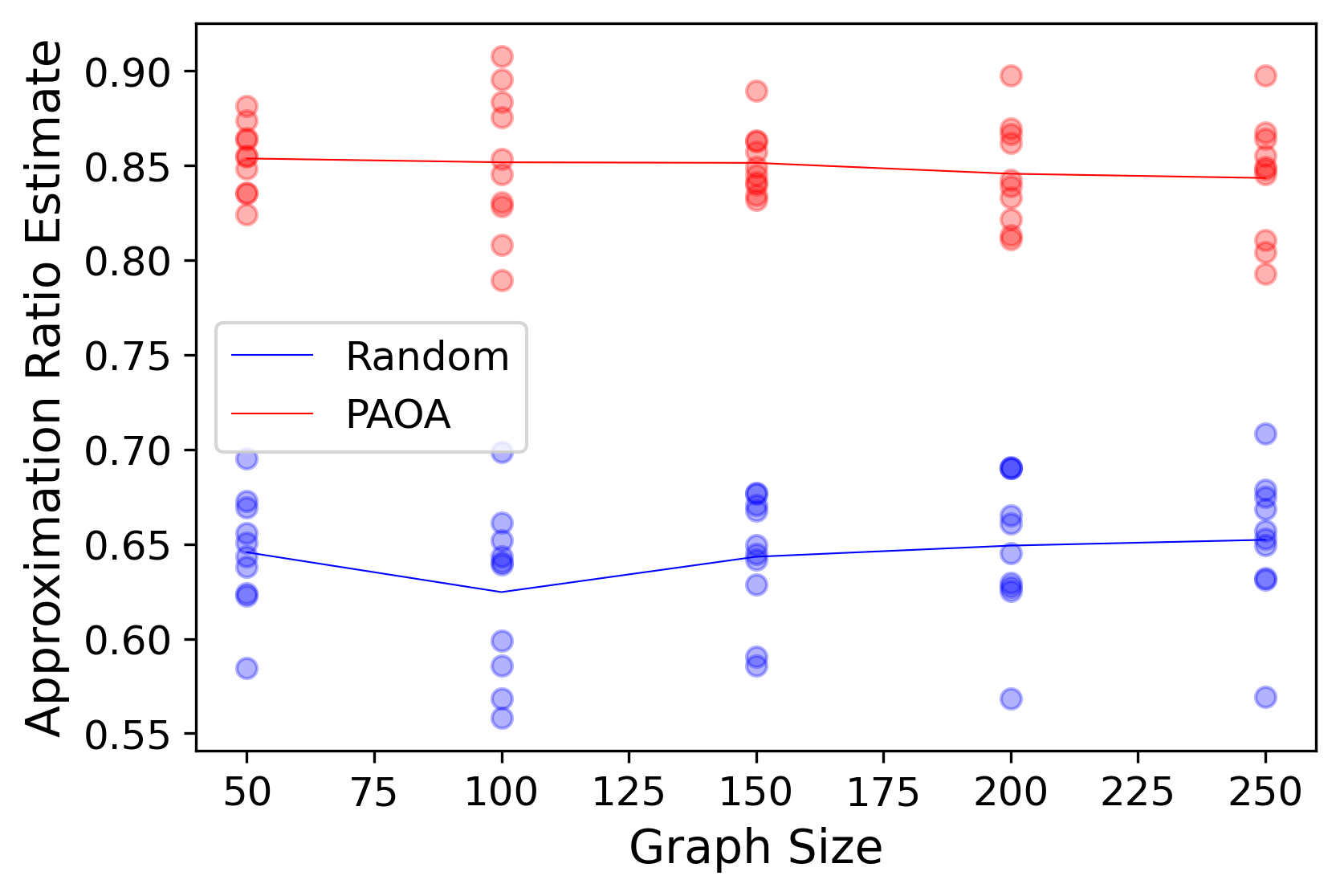}
\caption{A Comparison of Reduced PAOA and random guessing to solve max
cut on Barabasi-Albert graphs of increasing size. Each algorithm was run ten times on a Barabasi-Albert graph with a single initial node. All data are presented along with the average (line).}\label{fig:paoa_ratio_ba}
\end{figure}

As in Figure \ref{fig:paoa_ratio_2reg} from Section \ref{sec:reg_graphs}, we test the continued performance of PAOA on much larger graphs. In Figure \ref{fig:paoa_ratio_ba}, the results again demonstrate a reliable improvement over random guessing on a different class of random graphs. 

\section{CONCLUSION}\label{sec:conclusion}
In this work, we have introduced the idea of using sub-universal variational circuits to compare with quantum variational circuits. We explored a single-layer circuit composed of variational stochastic matrices, which can be physically realized. Numerical experiments have demonstrated that the ansatz we call the PAOA is preferred to the QAOA, its quantum analog. However, the following caveats apply. 

First and most obvious is the limited scope available to test quantum algorithms with classical simulators. It would be difficult to justify the extrapolation to practically relevant sized problems from such small qubit numbers. Second, PAOA and QAOA were trained using the same algorithm --- in other words, neither was optimized. The effect of this can be evidenced in some data not presented here for very small graphs. (The reader is encouraged to play with the simulations themselves at \cite{weitz_paoa_2023}). On small graphs, QAOA (as trained here) performs equally well as PAOA --- it is only on larger graphs that it struggles. We expect QAOA to perform better when its classical optimizer is optimized. Nevertheless, PAOA might also improve its performance with hyperparameter tuning.

The striking thing we have found is that PAOA works well ``out of the box.'' It is fast, convenient, and produces high-quality results with no tuning. If PAOA can be implemented directly in hardware (as in  \cite{2019pbits}, for example) and used to find solutions to CSPs in a cost-effective manner, it would impose a significant opposition to justifying investment in some quantum alternatives. In particular, the Reduced PAOA ansatz has good performance on a variety of graphs.  Based on the strong performance of Reduced PAOA on all the graphs considered, we conjecture that this improved performance is due to it finding the sweet spot where it has enough parameters to compute the cut but not too many to burden the optimizer. Moreover, as our ansatz is local, it would be interesting to prove if quantum methods have an advantage over classical as in Ref.~\cite{carlson_quantum_2023}.

Beyond the particular case we have solved, one could imagine benchmarking quantum machine learning protocols that have quantum inputs and quantum outputs in a similar fashion. In particular, one could construct parametric circuits out of other non-universal gate sets like (discrete) Cliffords~\cite{clifford1,clifford2} or Match gates~\cite{matchgate}. As the Match gates are continuously parameterized, it is easy to imagine using gradient descent type optimization, but the Cliffords could be optimized over genetic algorithms, for example.

In any case, we hope PAOA can provide a useful benchmark for testing parameterized quantum circuits in the future. While random guessing can provide a theoretical lower bound, PAOA serves as a practical performance gauge in numerical experiments.  

\noindent\textsc{Note:} Since our initial submission, we became aware of two related works. First, \citet{munoz-arias_low-depth_2023} show that an adaptively chosen depth-$N$ Clifford circuit can provide excellent approximate solutions to Max-Cut on a $N$-vertex graph. In contrast, our solution seems to find good approximate solutions with a depth-$1$ circuit and without adaption. Second, \citet{abdelrahman2025} extend our theoretical framework and realize PAOA experimentally. They developed hardware capable of executing PAOA, including the classical optimizer, on graphs with up to 500 vertices. This provides independent experimental validation of PAOA and a substantial generalization of our theoretical results.

{\em Acknowledgments:} The authors acknowledge helpful discussions with Charlie Carlson, Stuart Hadfield, Zackary Jorquera, Alexandra Kolla, Steven Kordonowy, Laurent Laborde, Nicholas Rubin, and Andrew Sornborger. LP was supported by the Sydney Quantum Academy, Sydney, NSW, Australia.

\bibliography{paoa.bib}

\end{document}